\documentclass[]{elsarticle}
\pdfoutput=1

\usepackage{amsmath}
\usepackage{amssymb}
\usepackage{float}
\usepackage{hyperref}
\usepackage{natbib}
\usepackage{pgf-umlcd}

\journal{Advances in Engineering Software}

\begin{document}
  
\begin{frontmatter}
\title{SplineLib: A Modern Multi-Purpose C++ Spline Library}

\author[cats]{Markus Frings}
\author[cats]{Norbert Hosters}
\author[cats]{Corinna M\"uller}
\author[cats]{Max Spahn}
\author[cats]{Christoph Susen}
\author[cats]{Konstantin Key\corref{corr1}}\ead{konstantin.key@cats.rwth-aachen.de}
\author[cats]{Stefanie Elgeti}
\cortext[corr1]{Corresponding author}
\address[cats]{Chair for Computational Analysis of Technical Systems\\Schinkelstra\ss{}e~2\\52062~Aachen\\Germany}

\begin{abstract}
  This paper provides the description of a novel, multi-purpose spline library.
  In accordance with the increasingly diverse modes of usage of splines, it is multi-purpose in the sense that it supports geometry representation, finite element analysis, and optimization.
  The library features reading and writing for various file formats and a wide range of spline manipulation algorithms.
  Further, a new efficient and objective-oriented algorithm for B-spline basis function evaluation is included.
  All features are available by a spline-type independent interface.
  The library is written in modern C++ with CMake as build system.
  This enables it for usage in typical scientific applications.
  It is provided as open-source library.
\end{abstract}

\end{frontmatter}
\textcopyright\;2020. This manuscript version is made available under the CC-BY-NC-ND 4.0 license \url{http://creativecommons.org/licenses/by-nc-nd/4.0/}

\section{Introduction}
Engineering design has to rely on a number of contributing aspects: innovation, designing, geometry representation, computing, planning, and predetermination of product properties \cite{Conrad2004}.
An important helper tool in this process is Computer Aided Design (CAD) software.
CAD is utilized to convey design information from the designer to the manufacturer.
As such, it has long ago replaced manual drawings.
While the way geometry is represented within CAD systems has remained unaltered for decades now, the role of the CAD software in the design process has indeed changed.
The trend certainly moves towards a more holistic design process, which includes also the steps of purchasing, manufacturing, and sales.
Furthermore---at least within the research community---CAD systems have become closely intertwined with analysis and optimization steps.

Given the importance of CAD software, we will now repeat the key ideas of geometry representation within this context.
A type of geometry representation common to most CAD systems are splines and in particular Non-Uniform Rational B-splines (NURBS).
As illustrated in \cite{Rogers2001}, the success of this spline type in CAD goes back to Versprille \cite{Versprille:1975aa} as well as Tiller and Piegl \cite{Tiller:1983aa,Piegl1987}.
As a member of the category of parametric geometry representations, NURBS can easily represent multi-valued functions.
Further, they feature a high level of smoothness and differentiability.
They are invariant under affine transformations and provide local control over geometry.
Finally, they are able to represent free-form shapes and---in the engineering context so important---conic sections \cite{Rogers2001}.
Common exchange formats for NURBS are IGES \cite{Smith:1986sm} or STEP \cite{pratt2001introduction}. 

In 2005, Hughes et.\ al introduced the concept of isogeometric analysis (IGA).
Here, the key idea is the use of spline functions as interpolation functions for the unknown solution in a finite element context.
Originally conceived to bridge the gap between CAD and analysis, the concept soon proved to be advantageous also solely from the analysis point of view.
As detailed in \cite{Cottrell2009}, key advantages of the IGA approach are: (1) exact (as in CAD-conforming) geometry representation; (2) user-controlled smoothness and continuity of the basis; (3) in general higher accuracy per degree of freedom; (4) in general variation diminishing property.
IGA has been applied to many different fields, including phase field methods for the Cahn-Hilliard equations \cite{Gomez2008}, brittle fracture \cite{Borden2012}, topology optimization \cite{Dede2012}, or fluid-structure-interaction \cite{Hsu2012,Hosters2018}. 

Another field of research that can profit largely from spline representations is shape optimization.
Here, a geometry representation based on splines not only ensures easy incorporation into the standard design and manufacturing process, but also provides a means to include shape constraints into the parameterization.
Shape optimization with splines in the form of free-form deformation is for example demonstrated in \cite{salmoiraghi2018free}, spline boundary representations are utilized in \cite{hirschler2018isogeometric,herrema2017framework,lian2016implementation,fusseder2015fundamental,kostas2015ship}.

In view of the requirements of CAD, IGA, and shape optimizations, the aim of this work is to propose an open source C++ spline library.
One question the reader might ask at this point is: ``Where is the need for yet another spline library?''.
Naturally, we are aware of the manifold spline libraries relying on a variety of programming languages and licensing concepts.
However, these libraries usually focus on one specific application.
In contrast, this library aims to be applicable to all three application areas mentioned above.

The predecessor of the proposed library was an in-house development written in FORTRAN.
It was essential in a variety of scientific publications, addressing shape optimization and IGA within the context of numerical flow simulations (e.g., \cite{siegbert2013design,siegbert2015individualized}).
However, it turned out that there is a high demand for a multi-purpose spline library in the scientific computing community.
Therefore, the development of a new C++ spline library was published as open source project under the LPGL3 license.
All experiences made while developing the FORTRAN library were used to improve the new development.

Besides a wide range of different functionality, also requirements regarding programming language, build system, and supported computer systems were taken into account.
The proposed spline library is fully written in modern C++, a programming language increasingly replacing FORTRAN in the scientific community.
We point out that it is easily possible to couple C++ code to other scientific programming languages like Python or FORTRAN.
With the intention of developing a state-of-the-art library, the most recent C++-17 standard is used.
As a result, the library can make use of modern object oriented designs; but the recent standard also requires a recent compiler.
However, almost every system can be equipped with a recent Clang, GCC, or Intel compiler.

As build system, CMake \cite{Kitware:2012} is used. This system is available for most high-performance computing (HPC) machines.
However, on HPC machines it is crucial to install software with all its dependencies.
The more dependencies a software package requires, the more complicated the installation becomes.
Fortunately, Gamblin et.\ al.\ \cite{Gamblin:2015aa} made a great effort in the last years to develop Spack as a dependency manager for scientific software.
The proposed software library comes with support for Spack.
Therefore, installation of the library and all its dependencies is quite simple.
That makes the software usable for a wide range of scientists without caring about dependencies.

To maintain a high software quality, SplineLib is tested by means of the Google testing framework \cite{Civil:2013}.
This is combined with continuous integration and a test coverage that is in the area of 100\%.
The resulting number of tests simultaneously serves as a set of examples on how to use the library.
There is no need to read a large documentation.
Instead, the user can look for a test case suitable to the application at hand.

The paper provides an overview over the main features and implementational constituents of SplineLib.
It concludes with two basic sample test cases from the area of IGA and shape optimization. 

\section{Splines}\label{splines}
This section gives a brief introduction into the definition of splines.
A detailed description of splines---as well as corresponding properties and algorithms---can be found in \cite{Piegl:2012aa}.

In general, splines map from a parametric space to a physical space.
Let there are $n+1$ basis functions $f_i$ defined in the parameter space.
In the \mbox{$N$-dimensional} physical space $X$, a control point $\mathbf{P}_i \in \mathbb{R}^N$ is associated with each basis function.
A parametric coordinate $\mathbf{u}$ is mapped to the physical space by (1) evaluating each basis function $f_i$ at $\mathbf{u}$; (2) multiplying these values with the corresponding control points; (3) summing up all products:
\begin{equation*}
  X\left(u\right) = \sum_{i = 0}^{n}{f_i\left(\mathbf{u}\right)\mathbf{P}_i}.
\end{equation*}

The definitions of the various spline types differ in the way the parametric space and the basis functions are defined.
SplineLib currently supports B-splines and the related NURBS, but implementation of further spline techniques is envisaged.
From the class of these splines, solely the so-called clamped splines are considered.
Clamped splines are those that interpolate their boundary control points exactly.

For simplicity, the current description is restricted to one-dimensional splines as higher dimensional splines (surfaces, volumes, etc.) are constructed by using tensor products of one-dimensional splines.
In contrast, the software library at hand supports splines of arbitrary dimension because the parametric dimensionality is implemented as a template parameter.

The parametric space is defined by a knot vector $U$.
A knot vector is a non-decreasing sequence of $m+1$ real numbers (potentially non-uniformly spaced):
\begin{equation*}
  U = \left\{u_0, \ldots, u_{m}\right\},
\end{equation*}
where $u_i \in \mathbb{R}$ is called a knot.
Based on the knot vector $U$, the $i$-th knot span is defined as:
\begin{equation}
  \begin{cases}
    \left[u_i, u_{i+1}\right), i \ne m - 1,\\
    \left[u_i, u_{i+1}\right], i = m - 1.
  \end{cases}
  \label{eq:knotspan}
\end{equation}
A knot span is called non-zero if $u_i \ne u_{i+1}$.
Given these knot spans, the B-spline basis functions are defined from $m$ basis functions of degree zero by using the Cox-de Boor recursion formula:
\begin{align}
  \begin{split}
    N_{i,0}\left(u\right) = &
    \begin{cases}
    1 & \text{if } u \in i\text{-th knot span},\\
    0 & \text{otherwise},
    \end{cases} \\
    N_{i,p}\left(u\right) = & \frac{u-u_i}{u_{i+p}-u_i}N_{i,p-1}\left(u\right) + \frac{u_{i+p+1}-u}{u_{i+p+1}-u_{i+1}}N_{i+1,p-1}\left(u\right).
  \end{split}
  \label{eq:bsplinebasisfnc}
\end{align}
The number of control points $n+1$ for a spline of degree $p$ is given as
\begin{equation*}
  n + 1 = m - p.
\end{equation*}
Note that the fraction $\frac{0}{0}$ can occur in Eq.\ \eqref{eq:bsplinebasisfnc} if a knot vector $U$ contains the same knot $p+1$ times.
This ratio is defined to be zero:
\begin{equation*}
  \frac{0}{0} := 0 \label{eq:divisionbyzero}.
\end{equation*}
Given these basis functions, a $p$-th degree B-spline curve can be written as
\begin{equation}
  \mathbf{C}\left(u\right) = \sum_{i=0}^nN_{i,p}\left(u\right)\mathbf{P}_i. \label{eq:bspline}
\end{equation}

It is well-known that B-splines have their limitations if geometries become more complex.
For example, it is not possible to exactly represent a circle by a B-spline curve.
NURBS overcome this limitation \cite{Versprille:1975aa,Tiller:1983aa}.
In contrast to B-splines, NURBS map to a weighted physical space.
Utilizing the B-spline basis functions, a NURBS curve is defined as
\begin{equation}
  \mathbf{C}\left(u\right) = \frac{\sum_{i=0}^nN_{i,p}\left(u\right)w_i\mathbf{P}_i}{\sum_{j=0}^nN_{j,p}\left(u\right)w_j},
  \label{eq:nurbsdefinition}
\end{equation}
where $w_i\in\mathbb{R}$ are the weights assigned to the control points.
Introducing
\begin{equation*}
  R_{i,p}\left(u\right) = \frac{N_{i,p}\left(u\right)w_i}{\sum_{j=0}^nN_{j,p}\left(u\right)w_j},
\end{equation*}
Eq.\ \eqref{eq:nurbsdefinition} can be rewritten as
\begin{equation*}
  \mathbf{C}\left(u\right) = \sum_{i=0}^nR_{i,p}\left(u\right)\mathbf{P}_i.
\end{equation*}
This form resembles the B-spline definition in Eq.\ \eqref{eq:bspline} with the exception of the definition of basis functions.

In the spline library both types of splines can be accessed by a common spline interface.
This allows for a spline-type independent code development.
Even new spline types as for example T-splines or PB-splines can be used without any relevant modifications if they are implemented within a later release of the software.

\section{Implementation: Evaluation of B-Spline Basis Functions}
There are manifold ways in which Eq.\ \eqref{eq:bsplinebasisfnc} can be translated into a computer algorithm.
Two key issues to keep in mind are possible repetitive evaluations of individual components and the occurrence of zero values in the denominator.

A commonly used and very efficient procedural algorithm for evaluating the B-spline basis functions is considered in \cite{Piegl:2012aa}.
In contrast to such procedural algorithms, the algorithm proposed here benefits from an object-oriented approach.
Its design is based on the idea of reflecting the evaluation procedure in a data structure.
Furthermore, it only checks once for zero denominators.
The key ideas are outlined in the next two sections.

\subsection{Data Structure for B-Spline Basis Functions} 
\begin{figure}
  \centering
  \scalebox{0.8}{
  \begin{tikzpicture}
    \begin{abstractclass}[text width = 11.4cm]{BasisFunction}{0, 1.75}
     	\attribute{- degree\_: Degree}
     	\attribute{- start\_knot\_: ParamCoord}
     	\attribute{- end\_knot\_: ParamCoord}
     	\attribute{- end\_knot\_is\_last\_knot\_: bool}
     	
     	\operation{+ Eval(pc: ParamCoord): double}
     	\operation{+ EvalDeriv(pc: ParamCoord, deriv: Derivative): double}
     	\operation{\# BasisFunction(kv: KnotVector, deg: Degree, start\_support: KnotSpan)}
     	\operation[0]{\# EvalOnSup(pc: ParamCoord): double}
     	\operation[0]{\# EvalDerivOnSup(pc: ParamCoord, deriv: Derivative): double}
     	\operation{- IsCoordInSup(pc: ParamCoord): bool}
    \end{abstractclass}
    
    \begin{class}[text width = 13.5cm]{BSplineBasisFunction}{0, -5}
     	\inherit{BasisFunction}
     	\attribute{- left\_denom\_inv\_: double}
     	\attribute{- right\_denom\_inv\_: double}
     	\attribute{- left\_lower\_degree\_: *BasisFunction}
     	\attribute{- right\_lower\_degree\_: *BasisFunction}
    
     	\operation{+ BSplineBasisFunction(kv: KnotVector, deg: Degree, start\_support: KnotSpan)}
     	\operation{\# SetLowerDegBasFncs(kv: KnotVector, deg: Degree, start\_support: KnotSpan): void}
     	\operation{\# EvalOnSup(pc: ParamCoord): double}
     	\operation{\# EvalDerivOnSup(pc: ParamCoord, deriv: Derivative): double}
     	\operation{- CompLeftQuot(pc: ParamCoord): double}
     	\operation{- CompRightQuot(pc: ParamCoord): double}
     	\operation{- InvWithPosZeroDenom(denom: double): double}
    \end{class}
    
    \begin{class}[text width = 13.5cm]{ZeroDegreeBSplineBasisFunction}{0, 5}
     	\inherit{BasisFunction}
     	\operation{+ ZeroDeZeroDegreeBSplineBasisFunction(kv: KnotVector, start\_support: KnotSpan)}
     	\operation{\# EvalOnSup (pc: ParamCoord): double}
     	\operation{\# EvalDerivOnSup (pc: ParamCoord, deriv: Deriv): double}
    \end{class}
    
    \begin{class}[text width = 14.5cm]{BasisFunctionFactory}{0, -11.25}
     	\operation{\underline{+ CreateDynamic(kv: KnotVector, start\_support: KnotSpan, deg: Degree): $\ast$BasisFunction}}
    \end{class}
  \end{tikzpicture}}
  \caption{UML class diagram for B-spline basis functions.} \label{fig:classDiagram}
\end{figure}
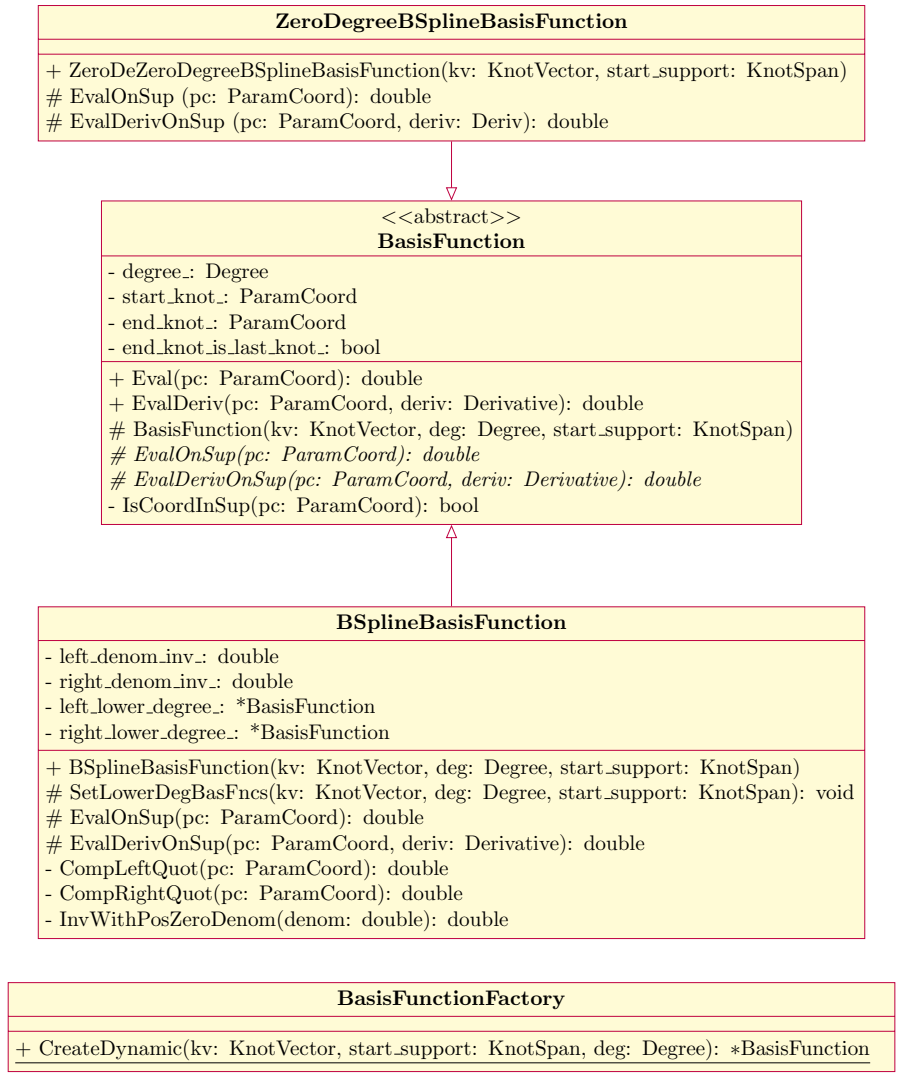

Reviewing the definition of the B-spline basis functions in Eq.\ \eqref{eq:bsplinebasisfnc} gives rise to the present design idea illustrated in Fig.\ \ref{fig:classDiagram}:
There are two types of B-spline basis functions $N_{i,p}$ that need to be distinguished, viz., of degree $p=0$ and of degree $p>0$.
The former are jump functions.
Instead, evaluating the $i$-th B-spline basis function of degree $p>0$ requires a linear combination of the two basis functions $N_{i,p-1}$ and $N_{i+1,p-1}$.

An abstract base class \texttt{BasisFunction} defines the common interface of B-spline basis functions of arbitrary degree $p\geq0$.
The basis function $N_{i,p}$ is characterized by its \texttt{degree\_} and its support specified by \texttt{start\_knot\_} and \texttt{end\_knot\_}.
The support starts with the \mbox{$i$-th} knot span (cf.\ Eq.\ \eqref{eq:knotspan}) and covers $p+1$ knot spans.
Therefore, it is usually given as $\left[u_i, u_{i+p+1}\right)$.
In the special case that $u_{i+p+1}$ equals the last knot $u_m$ of the knot vector $U$, $u_{i+p+1}$ must be included in the support (cf.\ second case of Eq. \eqref{eq:knotspan}).
Thus, the support is given as $\left[u_i, u_{i+p+1}\right]$ if \texttt{end\_knot\_is\_last\_knot\_} is set to \texttt{true} during the construction of an object of type \texttt{BasisFunction}.

Both \texttt{ZeroDegreeBSplineBasisFunction} and \texttt{BSplineBasisFunction} derive from this abstract class.
The former implements a step function whose support is a single knot span.
It represents the base case of the recursion.
The latter takes the linear combination in the Cox-de Boor recursion formula Eq. \eqref{eq:bsplinebasisfnc} into account  by storing pointers to objects of the basis functions of lower degree $N_{i,p-1}$ and $N_{i+1,p-1}$ in \texttt{left\_lower\_degree\_} and \texttt{right\_lower\_degree\_}, respectively.
As the associated denominators only depend on the knot vector, they do not change for a specific \texttt{BSplineBasisFunction}.
Hence, their adequately inverted values are computed during construction using \texttt{InvWithPosZeroDenom} and stored in \texttt{left\_denom\_inv\_} and \texttt{right\_denom\_inv\_}.

When the factory function \mbox{\texttt{CreateDynamic}} is called with a knot vector \texttt{kv}, a knot span \texttt{start\_support} (corresponding to the start of the basis function's support), and a degree \texttt{deg}, the object hierarchy is constructed.
An object of type \texttt{ZeroDegreeBSplineBasisFunction} or \texttt{BSplineBasisFunction} is created depending on whether \texttt{deg} is equal to zero or not.
Within the constructor of \texttt{BSplineBasisFunction}, the two objects for both lower degree B-spline basis functions are instantiated.

The B-spline basis functions' derivatives of arbitrary degree can also be computed by linearly combining corresponding derivatives of lower order B-spline basis functions.
Therefore, the proposed data structure is also well-suited for computing derivatives of arbitrary degree.  

\subsection{Evaluation Algorithm}
The methods \texttt{Eval} or \texttt{EvalDeriv} of a \texttt{BSplineBasisFunction} evaluate the function or its derivatives of arbitrary degree \texttt{deriv} at a parametric coordinate \texttt{pc}, respectively.
Within those two, \texttt{IsCoordInSup} is called to check if \texttt{pc} is contained in the support of the \texttt{BasisFunction} object.
If the parametric coordinate is not contained in the support, evaluating the functions and their derivatives yields zero.
Otherwise, \texttt{EvalOnSup} or \texttt{EvalDerivOnSup} are called.
Their actual implementations are provided by \texttt{ZeroDegreeBSplineBasisFunction} and \texttt{BSplineBasisFunction}.

Since \texttt{ZeroDegreeBSplineBasisFunction} is a jump function, it evaluates on its support simply to one while all derivatives are equal zero.
In contrast, evaluating a \texttt{BSplineBasisFunction} involves two non-trivial ingredients (cf.\ Eq.\ \eqref{eq:bsplinebasisfnc}): First, computing both fractions; second, evaluating the lower degree B-spline basis functions $N_{i,p-1}$ and $N_{i+1,p-1}$.
The two fractions are computed by calling \texttt{CompLeftQuot} and \texttt{CompRightQuot}.
Each method utilizes \texttt{start\_knot\_} or \texttt{end\_knot\_} to evaluate the numerator.
Next, its product with the values of the inverted denominator stored in \texttt{left\_denom\_inv} or \texttt{right\_denom\_inv} is computed.
These results are then multiplied with the values of the lower degree functions---accessed through \texttt{left\_lower\_degree\_} and \texttt{right\_lower\_degree\_}---and, finally, added.

\section{SplineLib Features}
This section describes the main utility functions of SplineLib: Input/Output and spline manipulation. Again, the multi-purpose idea has been emphasized.

\subsection{Input and Output}
SplineLib supports the four different file formats ITD, IGES, VTK and XML.
ITD is the native file format used in IRIT.
The application IRIT is a free-form geometric modeling environment developed by Elber \cite{Irit15}.
It supports modeling of rational and non-rational B-spline curves, surfaces, and trivariate volumes.
The latter is a feature almost unique to IRIT.
In the past, IRIT has been utilized in connection with IGA \cite{antolin2019isogeometric,antolin2019optimizing,massarwi2016b}.

A similar format to ITD is based on Extensible Markup Language (XML).
The XML-type data format for storing spline geometrical and related data was native to the FORTRAN library developed at the Chair for Computational Analysis of Technical Systems.
The XML format stores splines of arbitrary degree and dimension.
It was mainly developed in an IGA context.
Therefore, it can be used to store the spline geometry together with additional data.
The data can be defined as element data or control point data.
For practical reasons, it is only implemented for dimensions up to four.
This allows for typical three-dimensional computations as well as four-dimensional space-time computations, where time is handled as additional dimension.
However, SplineLib can easily be extended to support higher dimensions.
Parsing and writing XML files is built on top of the open-source library pugixml \cite{pugixml}.

Initial Graphics Exchange Specification (IGES) \cite{Smith:1983aa} is another file format that allows the digital exchange of geometry information among CAD systems.
The IGES format was introduced in 1979 and became the standard for transferring multi-dimensional models between CAD programs.
Although there are newer formats nowadays, it is still widely used.
Therefore, supporting IGES enables SplineLib to import and export data, that is relevant in industry.
IGES itself supports not only spline geometries, but also other types of data.
These are for example color information, meshes, and polylines.
The SplineLib implementation, however, focuses on NURBS and B-spline curves and surfaces.
All other IGES entries are ignored.

For post-processing data from computational fluid dynamics, an often used tool is Paraview \cite{Ahrens:2005aa}.
Paraview is an open-source software for data visualization and analysis.
It is built on top of the Visualization Toolkit (VTK) \cite{Schroeder:1999aa}.
VTK provides its own file format. Although it cannot represent splines, they can be visualized by converting them into a mesh.
As meshes cannot be transformed back into splines, SplineLib only provides a VTK writer and no reader.
This writer can be used to convert data obtained by SplineLib into mesh data for visualization and analysis.
The SplineLib API provides the user with the possibility to decide on the resolution of the mesh.
Thus, it can be adjusted to the relevant data.
The VTK export does not only convert the geometry, but can also add further element or control point data to the VTK file.
As a result, this exporter can be used to visualize data obtained by optimization or IGA with the well-developed tools for post-processing.

SplineLib also provides an interface to convert data between these different formats.
This functionality closes the gap between the different application areas.

\subsection{Spline Manipulation}
In addition to the support of multiple file formats, the second main feature implemented in SplineLib are the two most fundamental geometric algorithms for splines: (1) knot insertion/removal, and (2) subdivision.

Knot insertion is a means of improving the local control of the spline properties. Key to knot insertion is that it only acts on the parametric representation: It is important that the global geometry of the spline remains unaltered. In order to achieve this, new control points need to be inserted and others repositioned. The algorithm in SplineLib has been adapted from \cite{Piegl:2012aa}. In particular, the adaption concerns the spline dimension, which in SplineLib is implemented as template parameter.

In a similar fashion, knots can be removed.
Note that in contrast to knot insertion---depending on the specific scenario---knot removal may not be possible without altering the spline geometry.
In those cases, deviations (in terms of Euclidean distance) of the altered spline from the original one are bounded by a user-specified tolerance.

One particular extension of knot insertion is spline subdivision. It relies on inserting the same knot repeatedly, until $C^{-1}$-continuity is reached. At this point, the splines can be separated by splitting the original knot vector. Also this algorithm has been extended to arbitrary dimension.

\section{Application Examples}
This final section is devoted to two application examples: IGA and optimization.
\subsection{IGA}
\newcommand{\vekt}[1]{\mbox{\boldmath{${#1}$}}}
In order to demonstrate the applicability of SplineLib to solve partial differential equations (PDEs) using IGA, we consider Poisson's equation.
The general boundary value problem is given as:
\begin{equation*}
	-\Delta u = f \hspace*{0.1cm} on \hspace*{0.1cm} \Omega \hspace*{0.3cm} and \hspace*{0.3cm} \alpha u + \beta \frac{\partial u}{\partial n} = g \hspace*{0.1cm} on \hspace*{0.05cm} \partial \Omega, \medskip
\end{equation*}
with unknown function $u$, source term $f$, the computational domain $\Omega$ with boundary $\partial\Omega$ and constants $\alpha$, $\beta$, and some function $g$.
In the following, we restrict ourselves to homogeneous Dirichlet boundary conditions ($\alpha \ne 0, \beta = 0,$ and $g = 0$).
Furthermore, we assume a constant source term of $f = 1$.
The weak form is obtained by multiplication of the PDE with an arbitrary weighting function and integration over the computational domain $\Omega$.
Following Galerkin's method, the same function space is used for both the solution candidate and the weighting function.
The finite-dimensional subspace is represented by the spline basis function.
Considering the rational basis function $R$, the discrete weak form reads
\begin{equation*} 
	\sum_{j=0}^{n+1} \int_{\Omega} \nabla R_i \nabla R_j u_j \, d\Omega = \int_{\Omega} R_i f \, d\Omega \hspace*{0.5cm} \forall i = 0,\ldots,n+1. \label{eq:weak}
\end{equation*}
Note in particular that this approach uses the isoparametric principal, where the same basis functions are used for geometry representation and analysis.
Element-wise integration, i.e., integration on tensor-products of non-zero knot spans, by means of Gaussian quadrature then leads to
\begin{equation*}
	\sum_e \sum_k \sum_{\tilde{i}} \sum_{\tilde{j}} \nabla R_{\tilde{i}} \nabla R_{\tilde{j}} \, u_{\tilde{j}} \, w_k \, det(J_e) = \sum_e \sum_k \sum_{\tilde{i}} R_{\tilde{i}} \, f_k \, w_k \, det(J_e),
\end{equation*}
where $e$ denotes the element number, $k$ the integration points, $\tilde{i}$ and $\tilde{j}$ the indices of the non-zero basis functions and $w_k$ are the integration weights.
The resulting linear equation system is solved using \textit{Armadillo} \cite{sanderson2016armadillo,sanderson2018user}, an open-source linear algebra library.
More details on the formation of the linear equation system and IGA in general are provided in \cite{Hughes:2005}.

The IGA framework based on SplineLib distinguishes itself from other IGA codes by the implementation of the parametric dimensionality of the spline as template parameter.
Thereby, the user is enabled to solve PDEs on splines of arbitrary dimensionality using the same code.

\begin{figure}[H]
  \centering
  \includegraphics[width=0.7\linewidth]{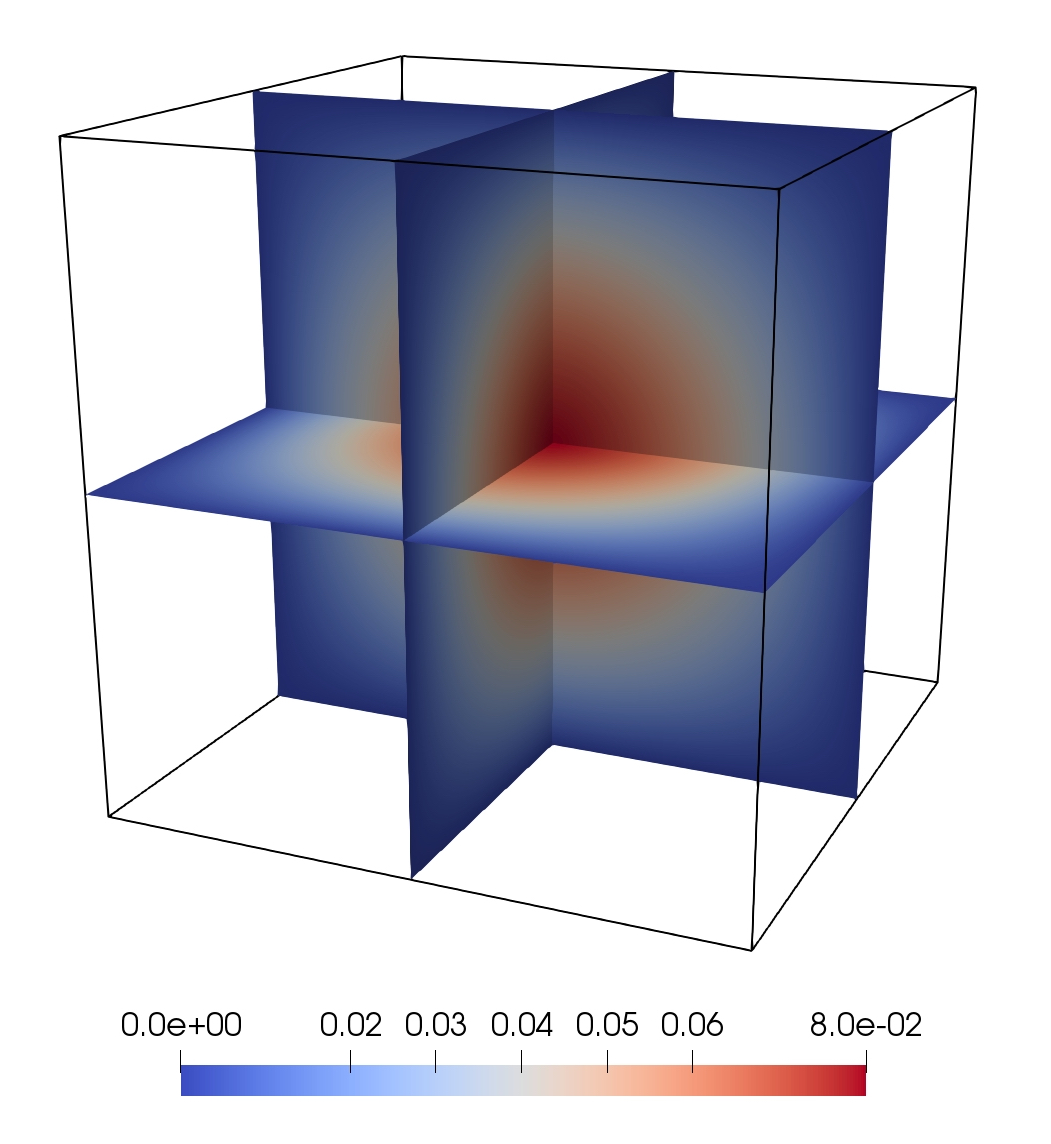}
  \caption{Solution of the reference problem on the unit cube.}
  \label{fig:iga_cube_1}
\end{figure}

The functionality of the IGA framework was verified using analytical solutions provided in \cite{elman} and solutions computed using finite element solvers.
To demonstrate the capability  to compute solutions for problems of higher dimensionality, the solution of the given reference problem is approximated inside the unit cube $[0, 1]^3$.
This problem is equivalent to finding the temperature distribution inside a cube with uniform heating and faces that are kept at zero temperature.
To visualize the solution, three slices through the cube are created normal to the $x$-, $y$- and $z$-axis.
These slices are shown in Fig.~\ref{fig:iga_cube_1}.
The temperature distribution is equal on all slices and is equal to the solution that would be obtained for the same boundary value problem on the square plate $[0, 1]^2$.
The highest temperature is reached in the center of the cube and decreases when getting closer to the boundary of the cube.
The temperature distribution is spherically symmetric. 

The IGA framework in SplineLib can certainly also be used for much more complex geometries.
A simple example is used here, since the exact solution of this problem can be represented using a series expansion.
This is not the case for complex geometries.

\subsection{Optimization}
A simple shape optimization problem is considered to demonstrate the value of SplineLib's functionalities.
The shape to optimize is a second-degree B-spline curve with the following properties:
The knot vector is given as
\begin{equation*}
  \{0.0, 0.0, 0.0, 1.0, 1.0, 1.0\}
\end{equation*}
and the shape is represented by the three two-dimensional control points
\begin{equation*}
  \{-1.0, 0.0\}, \{0.0, y\}, \{1.0, 0.0\},
\end{equation*}
where $y$ is the design variable for the optimization.
The design variable is subject to a bound constraint by $y \in [-1,1]$.
The initial value is $y=0$.
How to construct (as well as evaluate) this initial spline is shown in Fig.~\ref{fig:examplecode} (further examples can be found at https://github.com/mfcats/SplineLib/tree/master/example).
\begin{figure}
  \texttt{\footnotesize%
    array<Degree, 1> degree = $\lbrace$Degree$\lbrace$2$\rbrace\rbrace$;\newline
    KnotVectors<1> knot\_vector\_ptr =\vspace{-2.5mm}
    \begin{tabbing}
      \phantom{\hspace{5mm}}$\lbrace$make\_shared<KnotVector>(KnotVector($\lbrace$\=ParamCoord$\lbrace$0$\rbrace$, ParamCoord$\lbrace$0$\rbrace$,\\
      \>ParamCoord$\lbrace$0$\rbrace$, ParamCoord$\lbrace$1$\rbrace$,\\
      \>ParamCoord$\lbrace$1$\rbrace$, ParamCoord$\lbrace$1$\rbrace\rbrace$))$\rbrace$;
    \end{tabbing}\vspace{-6mm}
    \begin{tabbing}
      vector<ControlPoint> control\_points = $\lbrace$\=ControlPoint($\lbrace$-1.0, 0.0$\rbrace$),\\
      \>ControlPoint($\lbrace$ 0.0, 0.0$\rbrace$),\\
      \>ControlPoint($\lbrace$ 1.0, 0.0$\rbrace$)$\rbrace$;
    \end{tabbing}\vspace{-4.5mm}
    BSpline<1> b\_spline(knot\_vector\_ptr, degree, control\_points);\vspace{-.5mm}
    b\_spline.Evaluate($\lbrace$ParamCoord$\lbrace$0.5$\rbrace\rbrace$, $\lbrace$1$\rbrace$);
  }
  \caption{Initial spline constructed and evaluated in the second dimension at $u = 0.5$.}
  \label{fig:examplecode}
\end{figure}

The target shape is a parabola given as
\begin{equation*}
  y\left(x\right) = -x^2+1.
\end{equation*}
Thus, the objective function is the area between the B-spline and the parabola on the interval $y\in[-1,1]$.

To solve this simple optimization problem, an executable using SplineLib evaluates the objective function for a given control point coordinate $y$.
This objective function is coupled to the optimization framework Dakota \cite{Adams:2018aa}.
The gradient-free COBYLA algorithm is applied to the optimization problem.
For visualization purposes, VTK output is used.
The result of each optimization step is written into a single VTK file.

\begin{figure}[ht]
  \centering
  \includegraphics[width=0.5\linewidth]{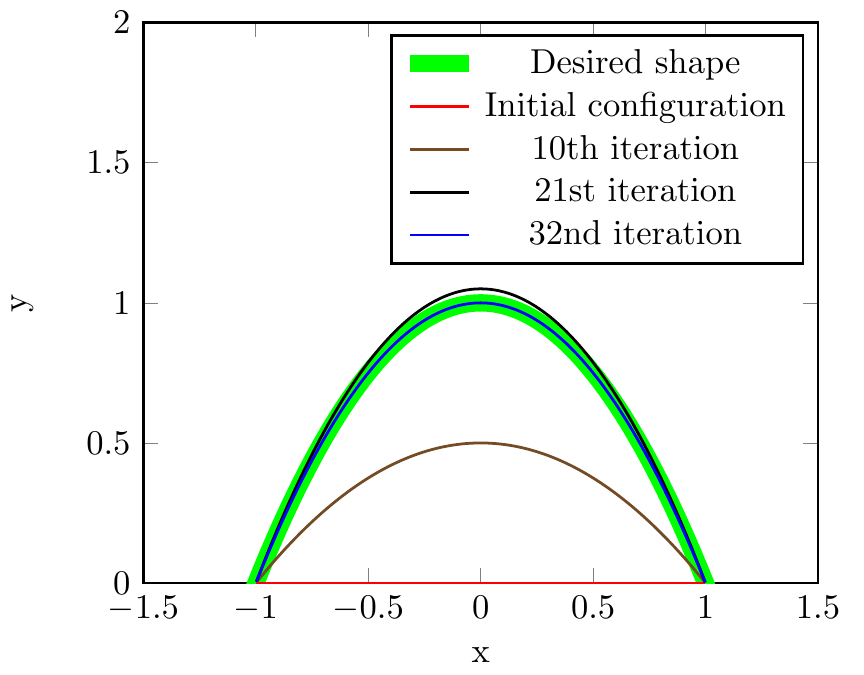}
  \caption{Solution steps for a simple shape optimization problem.}
  \label{fig:optimization_steps}
\end{figure}

The result is shown in Fig.~\ref{fig:optimization_steps}.
The solution converges to the target shape within 32 iterations.

With this example the application of SplineLib in an optimization environment is shown.
Due to the amount of supported input and output formats it is a straightforward step to use SplineLib also in more complex and real-world applications.

\section{Conclusion}
A new multipurpose C++ spline library is presented.
The aim of the library is to close the gap between different research fields.
Therefore, interfaces to commonly used file formats are provided.
With being able to represent not only curves and surfaces, but also splines of higher dimensions, SplineLib can be used in many applications.
For the evaluation of basis functions and their derivatives, a newly developed algorithm is facilitated.
In this algorithm, the procedural approach of evaluating the basis functions is realized as a data structure.
The main application areas SplineLib aims to connect are CAD, IGA, and optimization.
How to use SplineLib in these areas is shown in simple application examples.

From this initial state, it is necessary to further develop SplineLib.
It is required to make SplineLib more powerful, e.g., by supporting recent spline techniques and incorporating elaborate algorithms.
Here, we welcome the input of the community via GitHub (https://github.com/mfcats/SplineLib).

\section*{Acknowledgements}
\par {
  The authors gratefully acknowledge the support of Deutsche Forschungsgemeinschaft (DFG, German Research Foundation) under the Collaborative Research Center SFB 1120, subproject B2, the Cluster of Excellence ``Integrative Production Technology for High-Wage Countries'' EXC 128, and the International Research Training Group ``Modern Inverse Problems'' 33849990/GRK2379.
}

\section*{Bibliography}
\bibliographystyle{plain}
\bibliography{splinelib}
  
\end{document}